\documentclass[prd,aps,preprint,tightenlines,superscriptaddress]{revtex4}
\usepackage{epsfig}
\usepackage{amsmath}
\preprint{DOE/ER/40762-297} \preprint{UM-PP\#04-006}

\begin{document}
\title{The Beam Single Spin Asymmetry in Semi-inclusive Deep Inelastic Scattering}
\author{Feng Yuan}
\email{fyuan@physics.umd.edu}
\affiliation{Department of Physics,
University of Maryland,
College Park, Maryland 20742}
\begin{abstract}
We study the beam single spin asymmetry in semi-inclusive hadron
production in deep inelastic scattering in at order $1/Q$. There
are two competing contributions: the leading order transverse
momentum dependent parton distribution $h_1^\perp(x,k_\perp)$
convoluted with chiral-odd fragmentation function $\hat e(z)$, and
the chiral-odd distribution function $e(x)$ convoluted with
Collins fragmentation function $H_1^\perp(z,k'_\perp)$. We
estimate this asymmetry and compare with the experimental
measurements from CLAS and HERMES collaborations.
 \vspace{10cm}
\end{abstract}
\maketitle
\newcommand{\be}{\begin{equation}}
\newcommand{\ee}{\end{equation}}
\newcommand{\ben}{\[}
\newcommand{\een}{\]}
\newcommand{\beqn}{\begin{eqnarray}}
\newcommand{\eeqn}{\end{eqnarray}}
\newcommand{\Tr}{{\rm Tr} }


The single spin asymmetry (SSA) is a novel phenomena in high
energy spin physics, and has attracted much interest in recent
years \cite{review}. In particular, the measurements from the
HERMES, SMC, and JLAB collaborations show a remarkably large SSA
in the semi-inclusive deep inelastic scattering, such as pion
production in $\gamma^*p\rightarrow \pi X$, when the proton is
polarized transversely to the direction of the virtual photon
\cite{hermes,smc,jlab}. On the theoretical side, there are many
approaches to understanding SSA using Quantum Chromodynamics (QCD)
phenomenology \cite{review,qiu}. Recent interest focuses on the
transverse momentum dependent (TMD) parton distributions and their
implications to the semi-inclusive processes in deep inelastic
scattering
\cite{Collins:1981uk,sivers,Collins:1992kk,Ji:1993vw,Levelt:1994np,
mulders,boer,others}. For example, the Sivers function is one of
those TMD parton distributions representing the asymmetric
distribution of quarks in a transversely polarized proton, which
correlates the quark transverse momentum and the proton
polarization vector $\vec{S}_\perp$ \cite{sivers}. It contributes
to the target SSA in semi-inclusive deep inelastic scattering. The
existence of the Sivers function has been confirmed recently
\cite{brodsky,collins,ji1,mulders2}, where the final state
interactions from the gauge link in the gauge invariant definition
of TMD parton distributions play an important role.

In this paper, we will study the SSA related to the beam
polarization. Unlike the target SSA, the beam SSA is subleading in
$1/Q$, which will eventually vanish as $Q^2\to \infty$. However,
at some intermediate $Q^2$, but still large enough to guarantee
the factorization, this asymmetry might be important and
measurable. Experimentally, HERMES collaboration found this
asymmetry consistent with zero \cite{hermes}, but the CLAS
collaboration at JLab found sizable asymmetry, in the order of
$4\%$ \cite{Avakian:2003pk}. In the literature
\cite{Levelt:1994np,mulders,{Efremov:2002ut}}, this asymmetry has
been associated with the twist-3 chiral-odd distribution function
$e(x)$ \cite{Jaffe:1991kp} convoluted with the Collins
fragmentation function $H_1^\perp(z,k'_\perp)$
\cite{Collins:1992kk}. However, this is not the complete picture
at this order. There is an additional contribution: the leading
order TMD parton distribution $h_1^\perp(x,k_\perp)$ \cite{boer}
convoluted with a chiral-odd fragmentation function $\hat e(z)$
\cite{Ji:1993vw,Jaffe:1993xb}. We will demonstrate the existence
of this contribution below.

The TMD parton distribution $h_1^\perp$ represents the correlation
between the quark's transverse momentum and polarization in an
unpolarized proton state \cite{boer,Boer:1999mm}. It has the same
features as the Sivers function: it is a leading-order
distribution; it is nonvanishing due to the final state
interactions; it depends on the quark orbital angular momentum of
the nucleon \cite{ji3}. Various model calculations have also shown
that it has a similar size as the Sivers function
\cite{Gamberg:2003ey,Boer:2002ju,Yuan:2003wk}. On the other hand,
since $h_1^\perp$ is chiral-odd, it is very difficult to probe it
in deep inelastic scattering, just like the transversity
distribution. Our analysis shows that it contributes to the beam
SSA, which can be used to extract the distribution itself. Another
possible way to study $h_1^\perp$ distribution is the asymmetry in
the Drell-Yan process \cite{Boer:1999mm}.

We first derive the beam SSA in deep inelastic scattering. The
semi-inclusive hadron production cross section can be expressed as
\begin{equation}
\frac{d^5\sigma}{dx_Bdydzd^2P_{\perp
h}}=\frac{2\pi\alpha^2}{4zx_BQ^2s}L_{\mu\nu}W^{\mu\nu}\ ,
\label{xs}
\end{equation}
where $L_{\mu\nu}$ and $W^{\mu\nu}$ are leptonic and hadronic
tensors, respectively. We work in a frame where the virtual
photon's momentum $q$ and the proton's momentum $P$ are in the $z$
direction, and the incident and outgoing lepton's momenta $\vec l$
and $\vec l'$ form a scattering plane. We can define the azimuthal
angle of any momentum as an angle relative to the scattering
plane. The variable $s$ is the lepton-hadron total energy square,
$Q^2=-q^2$ the virtuality of the photon, and the dimensionless
variables $x_B$, $y$, and $z$ are defined as $x_B=Q^2/2P\cdot q$,
$y=P\cdot q/P\cdot l$, $z=P\cdot p_\pi/P\cdot q$. The variable
$P_{\perp h}$ is the transverse component of the observed pion's
momentum $p_\pi$. We further introduce two light-like vectors: $p$
and $n$, which satisfy $p^2=0$, $n^2=0$, $p\cdot n=1$, $p^-=0$,
and $n^+=0$. All momenta can be expressed in terms of $p$ and $n$
and the transverse momentum component. The leptonic tensor has the
symmetric and anti-symmetric parts,
$L_{\mu\nu}=-Q^2g_{\mu\nu}+2(l_\mu l'_\nu+l_\nu l'_\mu)+2
i\lambda_e\epsilon_{\mu\nu\rho\sigma}l^\rho l^{\prime \sigma}$,
where $\lambda_e$ is the polarization parameter of the lepton. The
antisymmetric part will give the beam spin asymmetry, convoluted
with the antisymmetric part of the hadronic tensor $W^{\mu\nu}_a$.

\begin{figure}[t]
\centerline{\psfig{figure=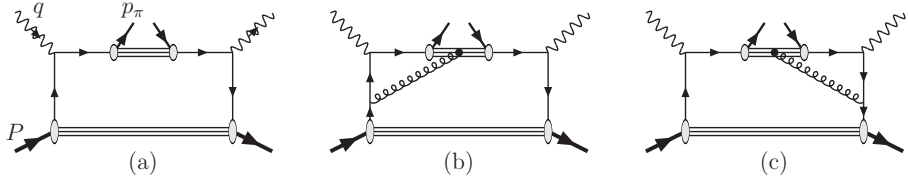,width=12 cm}} \caption{The
relevant diagrams contributing to the beam spin asymmetry
associated with $h_1^\perp\otimes \hat e(z)$ for semi-inclusive
pion production in deep inelastic scattering. }
\end{figure}

We follow the previous studies
\cite{Ji:1993vw,{Levelt:1994np},{mulders}}, and use QCD
factorization to calculate the hadronic tensor $W^{\mu\nu}$, which
can be separated into soft and hard parts. The hard part can be
calculated in perturbative QCD; and the soft parts are
nonperturbative and can be parameterized in Lorentz-invariant and
gauge-invariant distribution and fragmentation functions. The
relevant Feynman diagrams are shown in Fig.~1. The lower part of
these diagrams represent the parton distribution of the target,
and the upper part involves the fragmentation of a quark into a
pion. The symmetric part of $W^{\mu\nu}$ in the leading order has
a contribution from diagram (a). The antisymmetric part has two
contributions. Since the contribution from the $e(x)\otimes
H_1^\perp(z,k'_\perp)$ term has been calculated in
Ref.~\cite{Levelt:1994np}, in the following we will focus on the
contribution from the $h_1^\perp(x,k_\perp)\otimes \hat e(z)$
term. Because $\hat e(z)$ is a twist-3 fragmentation function, we
need to include the diagrams (b) and (c) of Fig.~1 to get an
electromagnetically gauge-invariant result.

The parton distributions can be defined from the following density
matrix \cite{Collins:1981uk},
\begin{eqnarray}
{\cal M}_p(x,k_\perp) = p^+
   \int {\frac{d\xi^-d^2\xi_\perp} {(2\pi)^3}}
        e^{-i(\xi^-k^+-\vec{\xi}_\perp\cdot \vec{k}_\perp)}
        \langle PS|\overline{\psi}(\xi^-,\xi_\perp)
     {\cal L}^\dagger {\cal L}
\psi(0)| PS\rangle \ , \label{density}
\end{eqnarray}
where ${\cal L}$ is the gauge link \cite{collins,ji1}. The density
matrix has the following expansion \cite{boer},
\begin{equation}
{\cal M}_p=\frac{1}{2}f_1(x,k_\perp)\!\! \not\! p+\frac{1}{2M}
h_1^\perp(x,k_\perp)\sigma^{\mu\nu}k_\mu p_\nu
    + \dots \ ,
\end{equation}
where $f_1$ is the usual unpolarized unintegrated parton
distribution, and both $f_1$ and $h_1^\perp$ are leading order in
twist counting.

Similarly, for the pion fragmentation functions, we can define the
density matrix as \cite{Ji:1993vw,{Levelt:1994np},mulders},
\begin{eqnarray}
{\cal M}_\pi (z,k'_\perp)= \frac{n^-}{2z}
   \int \frac{d\eta^+ d^2\eta_\perp} {(2\pi)^3}
        e^{-i(\eta^+k^{'-}-\vec\eta_\perp\cdot \vec k_\perp)}
        \langle 0|{\cal L}{\psi}(\eta^+,\eta_\perp)|\pi X\rangle
        \langle \pi X|
     \overline \psi(0){\cal L}^\dagger|0\rangle \ , \label{pid}
\end{eqnarray}
where $z=p_\pi\cdot p/k'\cdot p$, and $-z\vec k'_\perp$ is the
transverse momentum of pion relative to the quark's momentum. This
fragmentation matrix density has the expansion \cite{Ji:1993vw},
\begin{equation}
{\cal M}_\pi=\frac{1}{2}\hat f_1(z,k'_\perp)\!\!\not
\!n+\frac{1}{2}\frac{M_p}{p_\pi\cdot p}\hat e(z,k'_\perp)+\cdots \
, \label{ez}
\end{equation}
where $\hat f_1$ is the usual unpolarized unintegrated
fragmentation function; and $\hat e$ is the twist-3 chiral-odd
fragmentation function. There are two twist-3 and chiral-odd
fragmentation functions for the pion \cite{Jaffe:1993xb}, but we
only keep the one which contributes to beam SSA. As argued in
\cite{Ji:1993vw,Jaffe:1993xb}, instead of the pion mass, we put
the nucleon mass as the coefficient in front of $\hat
e(z,k'_\perp)$, because the pion mass vanishes in the chiral limit
but the density matrix ${\cal M}_\pi$ does not.

The contribution from Fig.~1(a) to the hadronic tensor
$W^{\mu\nu}$ can be calculated as,
\begin{equation}
W^{\mu\nu(a)}=2 z\int d^2k_\perp d^2k'_\perp \delta^{(2)}(P_{\perp
h}/z-k_\perp+k'_\perp){\rm tr}\left [{\cal M}_p\gamma^\mu{\cal
M}_\pi\gamma^\nu\right ] \ .
\end{equation}
Substituting the expansions of the density matrices ${\cal M}_p$
and ${\cal M}_\pi$, we have
\begin{eqnarray}
W^{\mu\nu(a)}&=&2 z \int d^2k_\perp d^2k'_\perp
\delta^{(2)}(P_{\perp
h}/z-k_\perp+k'_\perp)\left\{f_1(x,k_\perp)\hat f_1(z,k'_\perp)
\left[p^\mu
n^\nu+p^\nu n^\mu-g^{\mu\nu}\right]\right.\nonumber\\
&&~~~~\left. +ih_1^\perp(x,k_\perp)\hat
e(z,k'_\perp)\frac{1}{p_\pi\cdot p}\left[p^\mu k_\perp^\nu-p^\nu
k_\perp^\mu\right]\right\} \ ,
\end{eqnarray}
where the electric charge of the quark and the sum over all quark
flavor are implicitly assumed. The first term in the bracket is
the symmetric part of the tensor, and the second one
antisymmetric. The symmetric part itself is electromagnetic gauge
invariant, while the antisymmetric part is not, i.e., $q_\mu
W^{\mu\nu(a)}\ne 0$. However, after including the contributions
from diagrams (b) and (c) in Fig.~1, we can recover the
gauge-invariance \cite{Ji:1993vw}. This leads to the following
result for the antisymmetric part: $i2h_1^\perp(x,k_\perp)\hat
e(z,k'_\perp)/z /Q^2\left[T^\mu k_\perp^\nu-T^\nu
k_\perp^\mu\right]$, where $T^\mu=2xP^\mu+q^\mu$.

Including also the contributions from the convolution of $e(x)$
with the Collins function $H_1^\perp(z,k'_\perp)$
\cite{Levelt:1994np}, we get the complete result for the
antisymmetric part of the hadronic tensor at order $1/Q$,
\begin{eqnarray}
W^{\mu\nu}_a&=&2 z \int d^2k_\perp d^2k'_\perp
\delta^{(2)}(P_{\perp
h}/z-k_\perp+k'_\perp)\left\{ih_1^\perp(x,k_\perp)\frac{\hat
e(z,k'_\perp)}{z}\frac{2}{Q^2}\left[T^\mu k_\perp^\nu-T^\nu
k_\perp^\mu\right]\right.\nonumber\\
&&~~~~ \left. -ixe(x,k_\perp)
H_1^\perp(z,k'_\perp)\frac{2}{Q^2}\left[T^\mu k_\perp^{'\nu}-T^\nu
k_\perp^{'\mu}\right]\right\} \ ,
\end{eqnarray}
and it contributes to the beam SSA. We modified the definition of
the Collins function $H_1^\perp$ in \cite{Levelt:1994np,mulders}
by a factor of $M_p/m_\pi$ by the same argument we used in
Eq.~(\ref{ez}) for the fragmentation function $e(z)$. The
definition of $e(x)$ follows \cite{Jaffe:1991kp}. Since $T^\mu$ is
on order of $Q$, we see that the above antisymmetric part will
contribute to the cross section in the order of $1/Q$. That means
the beam SSA will be $1/Q$ suppressed, which is different from the
Sivers effect contribution to the target SSA being the leading
order effect.

Substituting the hadronic tensor into the differential cross
section formula Eq.~(\ref{xs}), we will get:
\begin{eqnarray}
\frac{d^5\sigma}{dx_Bdydzd^2P_{\perp
h}}&=&\frac{4\pi\alpha^2}{x_By^2s}\int d^2k_\perp
d^2k'_\perp\delta^{(2)}(\frac{P_{\perp
h}}{z}-k_\perp+k'_\perp)\left
\{(1-y+\frac{y^2}{2})f_1(x,k_\perp)\hat f_1(z,k'_\perp)\right.\nonumber\\
&&~~~+\lambda_e\frac{2y\sqrt{1-y}}{Q}h_1^\perp(x,k_\perp)|\vec
k_\perp|\frac{\hat
e(z,k_\perp')}{z}\sin\phi_k\nonumber\\
&&~~~\left.-\lambda_e\frac{2y\sqrt{1-y}}{Q}x
e(x,k_\perp)H_1^\perp(z,k'_\perp)|\vec k'_\perp| \sin
\phi_{k'}\right\} \ ,
\end{eqnarray}
where $\phi_k$ and $\phi_{k'}$ are the azimuthal angles of the
momenta $\vec{k}_\perp$ and $\vec k'_\perp$ relative to the
scattering plane, respectively. For example, we define $\sin
\phi_k=\vec l\times \vec l'\cdot \vec k_\perp/|\vec l\times \vec
l'||\vec k_\perp|$. Since only one of the two transverse momenta
$k_\perp$ and $k'_\perp$ is relevant for the $\sin\phi$ asymmetry,
we can integrate out the other without assuming the transverse
momentum dependence of the distribution and fragmentation
functions. After that, the $\phi_k$ or $\phi_{k'}$ dependence
leads to $\phi_h$ dependence, where $\phi_h$ is the azimuthal
angle of the observed hadron relative to the scattering plane:
$\sin \phi_h=\vec l\times \vec l'\cdot \vec P_{\perp h}/|\vec
l\times \vec l'||\vec P_{\perp h}|$. Finally,
\begin{eqnarray}
 \frac{d^5\sigma}{dx_Bdydzd^2P_{\perp
h}}&=&\frac{4\pi\alpha^2}{x_By^2s}\left\{(1-y+\frac{y^2}{2}) \int
d^2k_\perp\delta^{(2)}(P_{\perp h}-zk_\perp)
f_1(x,k_\perp)\hat f_1(z)\right.\nonumber\\
&& +\lambda_e\frac{2y\sqrt{1-y}}{Q}\int d^2k_\perp
\delta^{(2)}(P_{\perp h}-zk_\perp) h_1^\perp(x,k_\perp)|\vec
k_\perp|\frac{\hat
e(z)}{z}\sin\phi_h \label{beam}\\
&&\left.+\lambda_e\frac{2y\sqrt{1-y}}{Q}\int d^2k'_\perp
\delta^{(2)}(P_{\perp h}+zk'_\perp) x
e(x)z^2H_1^\perp(z,k'_\perp)|\vec k'_\perp| \sin \phi_h \right\}\
,\nonumber
\end{eqnarray}
where a sign has changed in the last term because $\vec k'_\perp$
and $\vec{P}_{\perp h}$ have opposite directions. The integrated
fragmentation functions are defined as $f_1(z)=z^2\int d^2k'_\perp
f_1(z,k'_\perp)$, the same for $\hat e(z)$, and the distribution
function $e(x)=\int d^2k_\perp e(x,k_\perp)$.

We can further simplify the differential cross section by
integrating out the transverse momentum $P_{\perp h}$ but keeping
the dependence on $\phi$,
\begin{eqnarray}
\frac{d^5\sigma}{dx_Bdydzd\phi}&=&\frac{2 \alpha^2}{x_By^2s}\left
\{(1-y+y^2/2)f_1(x)\hat f_1(z)\right.\nonumber\\
&&~~~+2\lambda_ey\sqrt{1-y}\frac{M_p}{Q}h_1^{\perp(1/2)}(x)\frac{\hat
e(z)}{z}\sin\phi\nonumber\\
&&~~~\left.+2\lambda_ey\sqrt{1-y}\frac{M_p}{Q}xe(x)H_1^{\perp(1/2)}(z)\sin\phi\right\}
\ ,
\end{eqnarray}
where the integrated parton distribution $f_1(x)=\int d^2k_\perp
f_1(x,k_\perp)$, $h_1^{\perp(1/2)}(x)=\int
d^2k_{\perp}|k_\perp|/M_ph_1^\perp(x,k_\perp)$, and fragmentation
$H_1^{\perp (1/2)}(z)=z^2\int
d^2k'_{\perp}|k'_\perp|/M_pH_1^\perp(z,k'_\perp)$. If we write the
differential cross section as $d\sigma\propto 1+A_y \sin\phi$,
\begin{equation}
A_y=\frac{\lambda_e\int dy dz dx_B
\frac{2y\sqrt{1-y}}{x_By^2}\frac{M_p}{Q}\left(h_1^{\perp(1/2)}(x)\frac{e(z)}{z}+xe(x)H_1^{\perp
(1/2)}(z)\right)} {\int dy dz
dx_B\frac{1-y+y^2/2}{x_By^2}f_1(x)\hat f_1(z)}\ . \label{ay}
\end{equation}
The $x_B$ and $z$ dependence of $A_y$ can also be similarly
calculated. We note that the two contributions have exactly the
same dependence on $y$, which makes it difficult to distinguish
them experimentally.

With known distribution and fragmentation functions, we can
predict the beam SSA. However, up to now, except for the
unpolarized quark distribution $f_1$ and fragmentation $\hat f_1$,
these functions can only be estimated in models. In addition, the
model calculations are not consistent at the present stage. For
example, controversial predictions exist for the Collins
fragmentation function $H_1^\perp$
\cite{Bacchetta:2003xn,{Gamberg:2003eg}}, and we have a wide range
of predictions for the leading-order TMD parton distribution
$h_1^\perp(x,k_\perp)$ from models
\cite{{Gamberg:2003ey},Boer:2002ju,{Yuan:2003wk}}. So, a reliable
model prediction is not possible at present. However, we can still
gain some insight for these functions by comparison with the
experimental data. For example, in \cite{Efremov:2002ut}, the beam
SSA has been interpreted as the result of the Collins effect, and
the experimental data were used to extract the distribution
function $e(x)$.

\begin{figure}[t]
\centerline{\psfig{figure=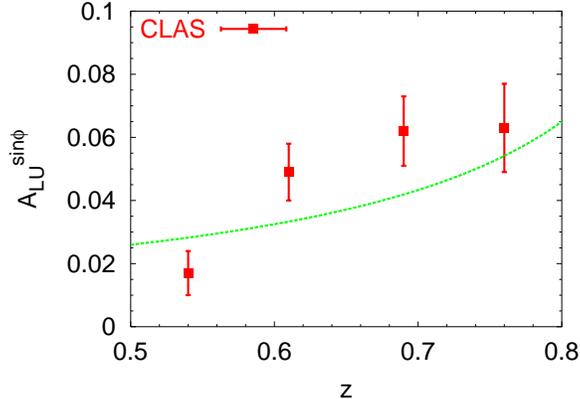,width=8cm}} \caption{The beam
SSA prediction from $h_1^\perp(x)\otimes \hat e(z)$, compared with
the experimental data from CLAS \cite{Avakian:2003pk}. The $z$
dependence solely comes from the ratio of the fragmentation
functions $\hat e(z)/f_1(z)$.}
\end{figure}

In this paper, we take an alternative extreme. We interpret the
beam SSA as a result of the first term in Eq.~(\ref{ay}). To
compare with the experimental data, we assume that the
factorization works at the energy range covered by the experiment.
This contribution depends on the chiral-odd fragmentation function
$\hat e(z)$, which has been calculated in a chiral quark model in
\cite{Ji:1993qx}. To a good approximation, we have
\begin{equation}
\hat e(z)=\frac{z}{1-z}\frac{m_q}{M_n}\hat f_1(z)\approx
\frac{1}{3}\frac{z}{1-z}\hat f_1(z)\ , \label{eze}
\end{equation}
where $m_q$ is the constituent quark mass, and $M_n$ the nucleon
mass. The above relation is only true at the scale of
$\Lambda_\chi$, and at higher scale their relation might breakdown
because the evolution of these two functions is different.
However, as a rough estimate, we will adopt such approximations.
The chiral quark model prediction for the usual unpolarized
fragmentation function $\hat f_1(z)$ is consistent with the
experimental data after considering the evolution effects
\cite{Ji:1993qx}. We should also note that the chiral quark model
is not suitable for the calculation of the fragmentation function
at $z\to 1$ region, where the invariant mass of the fragmenting
quark exceeds the cutoff of the model, $\Lambda_\chi$.

The $z$ dependence of the asymmetry $A_y$ only comes from the
ratio of the two fragmentation functions $\hat e(z)$ and $\hat
f_1(z)$ in Eq.~(\ref{ay}), and the simple relation
Eq.~({\ref{eze}) can be used to predict the $z$ dependence of
$A_y$. In Fig.~2, we show the normalized asymmetry prediction from
this term compared with CLAS measurements. The most striking
observation is that this simple relation Eq.~(\ref{eze}) agrees
with the experiment very well. The normalization of $A_y$ also
depends on the TMD parton distribution $h_1^\perp$. This
distribution involves more complicated dynamics
\cite{Gamberg:2003ey,{Boer:2002ju},{Yuan:2003wk}}, and hence is
less reliable compared to the fragmentation function $\hat e(z)$
in Eq.~(\ref{eze}). Nevertheless, from what we have now for
$h_1^\perp$, we can make an order-of-magnitude estimate and
compare with experiment. For example, a bag model calculation
shows that the ratio of $h_1^{\perp(1/2)}(x)/f_1(x)$ at the
kinematic region of the CLAS measurement $0.15<x<0.4$ is about
$0.04$ for $u$ quark \cite{Yuan:2003wk}. After taking into account
other kinematic factors in Eq.~(\ref{ay}), the asymmetry $A_y$ is
predicted to be about $0.05$, in rough agreement with the CLAS
result of $0.038$ \cite{Avakian:2003pk}, although the bag model
prediction is very crude and the sign is inconsistent.

Extending the above estimate to the HERMES kinematics, the beam
SSA is at least a factor of 5 less than what CLAS has found,
consistent with the HERMES measurement \cite{hermes}. This is just
the consequence of the beam SSA being $1/Q$ effect. So, there is
no contradiction between these two experiments. This also agrees
with the observation of \cite{Efremov:2002ut}.

We note that another interpretation of the beam SSA has been made
in \cite{Afanasev:2003ze}, where the photon ``Sivers" effect was
considered. In Ref.~\cite{Hagiwara:1982cq}, an ${\cal
O}(\alpha_s^2)$ QCD effect to the beam SSA have also been
investigated. We did not include these effects in our formalism.

In conclusion, we have calculated the beam single spin asymmetry
in semi-inclusive hadron production in deep inelastic scattering.
Up to $1/Q$, there are two contributions: the distribution
$h_1^\perp(x,k_\perp)$ convoluted with fragmentation $\hat e(z)$,
and $e(x)$ convoluted with $H_1^\perp(z,k'_\perp)$. A simple
chiral quark model prediction of $\hat e(z)\approx z/3/(1-z)\hat
f_1(z)$ agrees well with the experimental data on the $z$
dependence of the asymmetry. Further experimental data can provide
more information on the extraction of the leading order TMD parton
distribution $h_1^\perp$.

We thank H. Avakian and X. Ji for numerous discussion associated
with the topic of this paper. The author also thanks A. Bacchetta,
D. Boer, C. Carlson, and L. Gamberg for useful conversations. This
work was supported by the U. S. Department of Energy via grants
DE-FG02-93ER-40762.


\begin{thebibliography}
\frenchspacing
\bibitem{review} see, for example, reviews,
M.~Anselmino, A.~Efremov and E.~Leader,
Phys.\ Rept.\  {\bf 261}, 1 (1995) [Erratum-ibid.\  {\bf 281}, 399
(1997)];
V.~Barone, A.~Drago and P.~G.~Ratcliffe,
Phys.\ Rept.\  {\bf 359}, 1 (2002).

\bibitem{hermes}
A.~Airapetian {\it et al.}  [HERMES Collaboration],
Phys.\ Rev.\ Lett.\  {\bf 84}, 4047 (2000);
A.~Airapetian {\it et al.}  [HERMES Collaboration],
Phys.\ Rev.\ D {\bf 64}, 097101 (2001).

\bibitem{smc}
D.~Adams {\it et al.}  [Spin Muon Collaboration (SMC)],
Phys.\ Lett.\ B {\bf 336}, 125 (1994);
A.~Bravar  [Spin Muon Collaboration],
Nucl.\ Phys.\ A {\bf 666}, 314 (2000).

\bibitem{jlab}
H.~Avakian  [CLAS Collaboration], proceedings of ``Testing QCD
Through SPIN Observables" (Ed.D.G.Crabb et al.) University of
Virginia April 2002 (2002).

\bibitem{qiu}
J.~Qiu and G.~Sterman,
Phys.\ Rev.\ Lett.\  {\bf 67}, 2264 (1991);
Phys.\ Rev.\ D {\bf 59}, 014004 (1999).

\bibitem{Collins:1981uk}
J.~C.~Collins and D.~E.~Soper,
Nucl.\ Phys.\ B {\bf 193}, 381 (1981) [Erratum-ibid.\ B {\bf 213},
545 (1983)];
J.~C.~Collins and D.~E.~Soper,
Nucl.\ Phys.\ B {\bf 194}, 445 (1982).

\bibitem{sivers}
D.~W.~Sivers,
Phys.\ Rev.\ D {\bf 41}, 83 (1990)
[Annals Phys.\  {\bf 198}, 371 (1990)];
D.~W.~Sivers,
Phys.\ Rev.\ D {\bf 43}, 261 (1991).

\bibitem{Collins:1992kk}
J.~C.~Collins,
Nucl.\ Phys.\ B {\bf 396}, 161 (1993).

\bibitem{Ji:1993vw}
X.~Ji,
Phys.\ Rev.\ D {\bf 49}, 114 (1994).

\bibitem{Levelt:1994np}
J.~Levelt and P.~J.~Mulders,
Phys.\ Lett.\ B {\bf 338}, 357 (1994).

\bibitem{mulders}
P.~J.~Mulders and R.~D.~Tangerman,
Nucl.\ Phys.\ B {\bf 461}, 197 (1996) [Erratum-ibid.\ B {\bf 484},
538 (1997)].

\bibitem{boer}
D.~Boer and P.~J.~Mulders,
Phys.\ Rev.\ D {\bf 57}, 5780 (1998).

\bibitem{others}
M.~Anselmino, M.~Boglione and F.~Murgia,
Phys.\ Lett.\ B {\bf 362}, 164 (1995);
M.~Anselmino and F.~Murgia,
Phys.\ Lett.\ B {\bf 442}, 470 (1998).

\bibitem{brodsky}
S.~J.~Brodsky, D.~S.~Hwang and I.~Schmidt,
Phys.\ Lett.\ B {\bf 530}, 99 (2002);
Nucl.\ Phys.\ B {\bf 642}, 344 (2002).

\bibitem{collins}
J.~C.~Collins,
Phys.\ Lett.\ B {\bf 536}, 43 (2002).

\bibitem{ji1}
X.~Ji and F.~Yuan,
Phys.\ Lett.\ B {\bf 543}, 66 (2002);
A.~V.~Belitsky, X.~Ji and F.~Yuan,
Nucl.\ Phys.\ B {\bf 656}, 165 (2003).

\bibitem{mulders2}
D.~Boer, P.~J.~Mulders and F.~Pijlman,
Nucl.\ Phys.\ B {\bf 667}, 201 (2003).


\bibitem{Avakian:2003pk}
H.~Avakian {\it et al.}  [CLAS Collaboration],
arXiv:hep-ex/0301005.

\bibitem{Efremov:2002ut}
A.~V.~Efremov, K.~Goeke and P.~Schweitzer,
Phys.\ Rev.\ D {\bf 67}, 114014 (2003).

\bibitem{Jaffe:1991kp}
R.~L.~Jaffe and X.~Ji,
Phys.\ Rev.\ Lett.\  {\bf 67}, 552 (1991);
Nucl.\ Phys.\ B {\bf 375}, 527 (1992).

\bibitem{Jaffe:1993xb}
R.~L.~Jaffe and X.~Ji,
Phys.\ Rev.\ Lett.\  {\bf 71}, 2547 (1993).

\bibitem{Boer:1999mm}
D.~Boer,
Phys.\ Rev.\ D {\bf 60}, 014012 (1999).

\bibitem{ji3}
X.~Ji, J.~P.~Ma and F.~Yuan,
Nucl.\ Phys.\ B {\bf 652}, 383 (2003).

\bibitem{Gamberg:2003ey}
L.~P.~Gamberg, G.~R.~Goldstein and K.~A.~Oganessyan,
Phys.\ Rev.\ D {\bf 67}, 071504 (2003).

\bibitem{Boer:2002ju}
D.~Boer, S.~J.~Brodsky and D.~S.~Hwang,
Phys.\ Rev.\ D {\bf 67}, 054003 (2003).

\bibitem{Yuan:2003wk}
F.~Yuan,
arXiv:hep-ph/0308157.

\bibitem{Bacchetta:2003xn}
A.~Bacchetta, R.~Kundu, A.~Metz and P.~J.~Mulders,
Phys.\ Lett.\ B {\bf 506}, 155 (2001);
A.~Bacchetta, A.~Metz and J.~J.~Yang,
arXiv:hep-ph/0307282.

\bibitem{Gamberg:2003eg}
L.~P.~Gamberg, G.~R.~Goldstein and K.~A.~Oganessyan,
Phys.\ Rev.\ D {\bf 68}, 051501 (2003).

\bibitem{Ji:1993qx}
X.~Ji and Z.~K.~Zhu,
arXiv:hep-ph/9402303.

\bibitem{Hagiwara:1982cq}
K.~Hagiwara, K.~i.~Hikasa and N.~Kai,
Phys.\ Rev.\ D {\bf 27}, 84 (1983).

\bibitem{Afanasev:2003ze}
A.~Afanasev and C.~E.~Carlson,
arXiv:hep-ph/0308163.


\end{thebibliography}
\end{document}